\documentclass[aps,prl,twocolumn,longbibliography,superscriptaddress]{revtex4-2}
\usepackage[english]{babel}
\usepackage{bm}

\usepackage[a4paper,top=2cm,bottom=2cm,left=3cm,right=3cm,marginparwidth=1.75cm]{geometry}

\usepackage{textgreek}
\usepackage{amsmath}
\usepackage{graphicx}
\usepackage[colorlinks=true, allcolors=blue]{hyperref}

\newcommand{\I}{\mathrm{i}}
\newcommand{\be}{\begin{equation}}
\newcommand{\ee}{\end{equation}}

\usepackage{xcolor}
\definecolor{darkgreen}{rgb}{0, 0.45, 0}
\newcommand{\revisedtext}[1]{\textcolor{black}{#1}}
\usepackage{soul}

\begin{document}

\title{Ultrafast artificial intelligence: Machine learning with atomic-scale quantum systems}
\author{Thomas Pfeifer}
\email{thomas.pfeifer@mpi-hd.mpg.de}
\address{Max-Planck-Institut für Kernphysik, 69117 Heidelberg, Germany}
\author{Matthias Wollenhaupt}
\email{matthias.wollenhaupt@uni-oldenburg.de}
\address{Carl-von-Ossietzky Universit\"at Oldenburg, Institut f\"ur Physik, Carl-von-Ossietzky-Stra{\ss}e 9-11, 26129 Oldenburg, Germany}
\author{Manfred Lein}
\email{lein@itp.uni-hannover.de}
\address{Leibniz Universit\"at Hannover, Institut f\"ur Theoretische Physik, Appelstra{\ss}e 2, 30167 Hannover, Germany}



\date{\today}

\begin{abstract}

We train a model atom to recognize hand-written digits in the range 0-9, employing intense light--matter interaction as a computational resource. For training, the images of the digits  are converted into shaped laser pulses (data input pulses). Simultaneously with an input pulse, another shaped pulse (program pulse), polarized in the orthogonal direction, is applied to the atom and the system evolves quantum mechanically according to the time-dependent Schr\"odinger equation. The purpose of the optimal program pulse is to direct the system into specific atomic final states (classification states) that correspond to the input digits.  
\revisedtext{A success rate of about 40\% is achieved when using a basic optimization scheme that might be limited by the computational resources for finding the optimal program pulse in a high-dimensional search space.
%
Our key result is the capability of the laser-programmed atom to generalize, i.e.,~the classification of unseen images is improved by training.}  This atom-sized machine-learning image-recognition scheme  operates on time scales down to tens of femtoseconds, is scalable towards larger (e.g. molecular) systems, and  is readily reprogrammable towards other learning/classification tasks. 

\end{abstract}
\maketitle

Artificial intelligence (AI) is an area of growing interest and with an enormous range of applications, due to recent successes and breakthroughs in deep learning, enabled by the steady increase in (classical) computational power~\cite{Goodfellow-et-al-2016,ML}. Within this field, image recognition, i.e., the classification of different but conceptually equivalent (input) images into unique (output) categories, has been one of the prime applications of artificial intelligence and machine learning for many years~\cite{Plamondon2000}.  
\revisedtext{A key component of machine learning is the ability of the trained system to generalize \cite{Vapnik1998,Neyshabur2017, Kawaguchi2022generalization}, i.e., to correctly classify input data that was not part of the training data.}

With recent advances in optical science and technology, in particular optical neuromorphic hardware~\cite{Feldmann2019}, it has recently been possible to accelerate image recognition to sub-nanosecond timescales~\cite{Ashtiani2022}. Here, the operation timescale is determined by the speed of light at which information propagates in microscopic waveguides on the chip, and is thus directly proportional to its size:  The smaller the device and the computational units, the faster the clock rates for the recognition operations will become.  The fundamental limit to further minimization is the atomic scale.

In atomic and molecular physics, extreme timescales down to femtoseconds and even attoseconds are currently being explored and controlled by measuring and steering the motion of one, two, or more electrons~\cite{Weinacht1999,Ott2014, Pengel2017,Jiang2022,Kretschmar2022,Yu2022} with lasers. Atomic states are excited and coupled on ultrafast (femtosecond) time scales~\cite{Meister2021}, providing a quantum analogue of neurons (quantum states) and axons (laser coupling between states). Ground-state atoms and molecules are fully quantum-correlated systems and therefore entanglement arises naturally when these systems fragment into two or more particles~\cite{Akoury2007,Schoffler2008}, e.g. due to photoionization or dissociation. Control of entangled states using attosecond and femtosecond laser pulses has also been addressed recently~\cite{Vrakking2021,Koll2022,Busto2022,Shobeiry2022}. The natural question is whether atoms and their interaction with intense laser light can be used as a high-speed computational resource in applications such as machine learning for image recognition. This question has recently been addressed for the case of two-class recognition of hand-written digits and three-class recognition of iris-flower types, using the process of high-order harmonic generation and thus an optical output channel~\cite{McCaul2023}.  

\begin{figure}
    \includegraphics[width=\columnwidth] {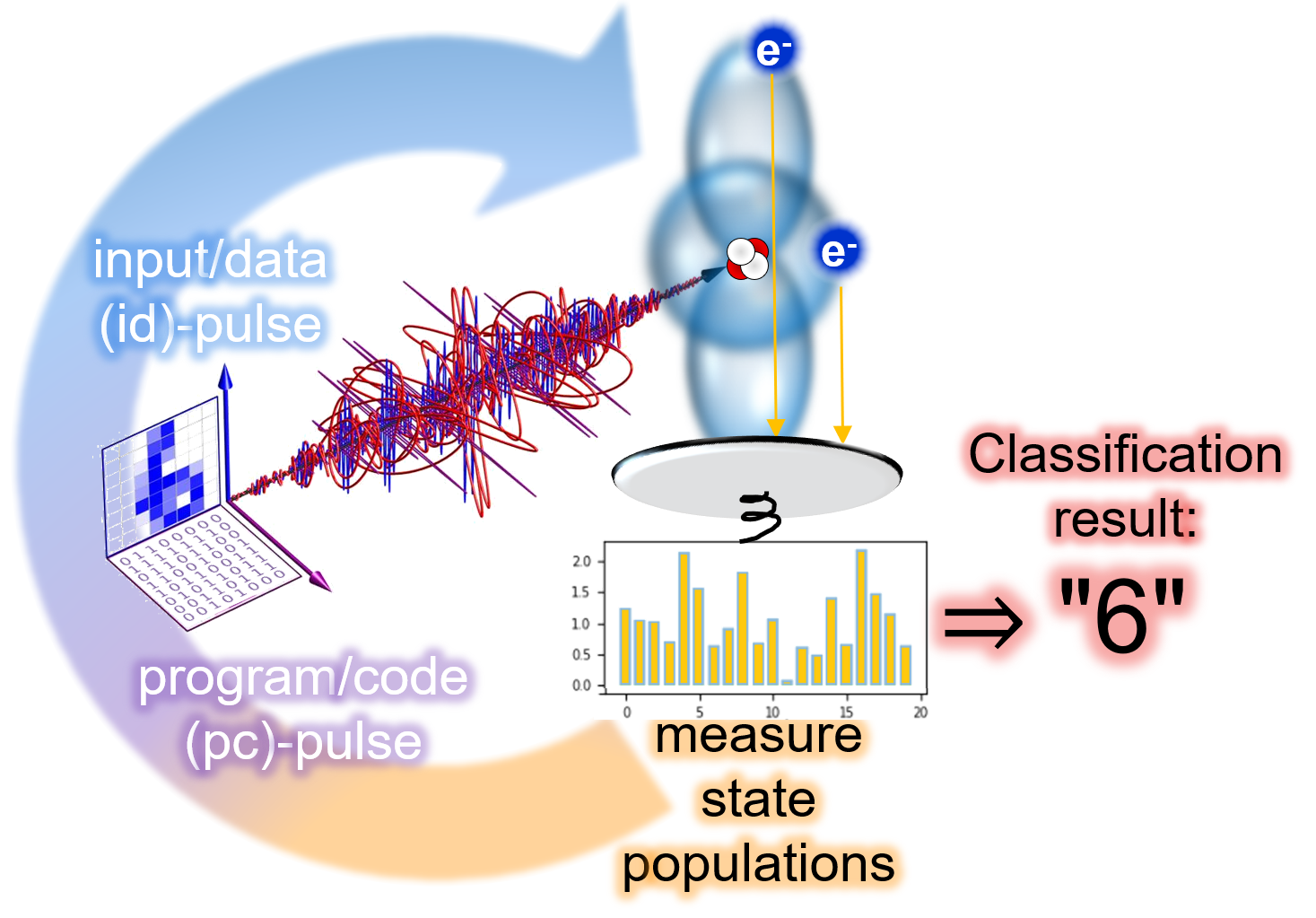}
    \caption{\label{ConceptualIdea} Conceptual representation of atomic machine learning. A quantum system (here: an atom) interacts with an input/data (id) pulse and a program/code (pc) pulse (here: along two orthogonal polarization directions) to deliver the data and the code, respectively.  The quantum state populations after the interaction are read out (e.g. by projecting them into the continuum and employing an electron spectrometer) and the maximum population among the  classification states (after renormalization) indicates the classification result.  The pc pulse is found by training on a large number of hand-written digits.}
\end{figure}

In the present work, we investigate whether an atom could at the same time act as a quantum processor \emph{and} readout register for machine learning, mapping two-dimensional images of digits directly onto atomic quantum states, referred to as classification states in the following. The latter could then be either read out or serve as input to subsequent (quantum) processing tasks. Our approach is to be distinguished from recent developments referred to as {quantum machine learning}~\cite{Biamonte2017,Beer2020}, where the ultimate goal is to train quantum computing devices to perform quantum tasks, and also from the earlier proposals to mimic specific quantum gates in laser-driven molecules~\cite{Tesch2002,Palao2002,Teranishi2006} or Rydberg atomic systems~\cite{Ahn2000}. In our case, we train quantum systems directly \revisedtext{and demonstrate generalization capability}, without the intention of developing a quantum algorithm as sequence of established quantum gates.
\revisedtext{
Being based on the time evolution of a quantum system, our scheme benefits from the same quantum effects that can enable quantum speedup in the framework of quantum computing, see the discussions specifically in the context of classification tasks \cite{Liu2021,Sharma2023}.
}

In the following, we introduce our scheme and present the results of a model-atom simulation as a proof of principle. Both the input data and the code that processes the data are supplied to the atom in the form of shaped femtosecond light pulses, namely input/data(id)~pulse and program/code(pc)~pulse (see Fig.~\ref{ConceptualIdea}). Due to the ultrafast time scale, incoherent coupling to the environment can be neglected, i.e., the dynamics of the atom is described by the time evolution of a pure state evolving according to the time-dependent Schr\"odinger equation (TDSE) 
$\I \partial_t\Psi=\mathcal{H}\Psi$ (atomic units are used unless stated otherwise). We show that the atom can be successfully trained, despite a huge parameter space for the optical pc pulse, which is obtained by an evolutionary optimization procedure~\cite{Judson1992}. We expect that the training approach will be significantly improved in the future, e.g. by using more suitable parameterizations for the program pulse and advanced statistical methods based on Bayesian inference.

\begin{figure}
    \includegraphics[width=\columnwidth] {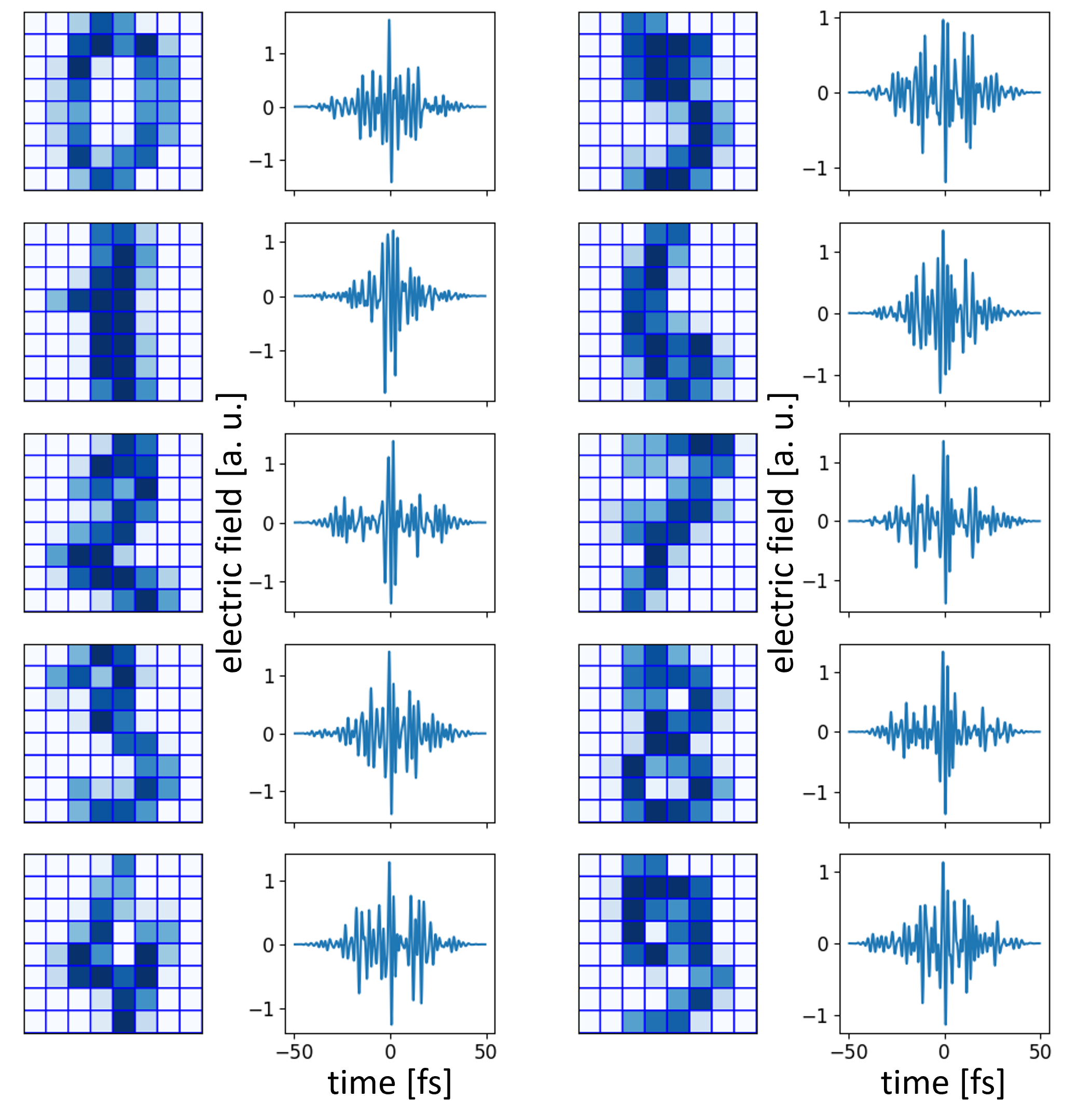}
    \caption{\label{fig:PulseShapes} A sample of hand-written digits and their encoding into electric fields (in arbitrary units) along one of the polarization axes.}
\end{figure}

\begin{figure}
    \includegraphics[width=\columnwidth] {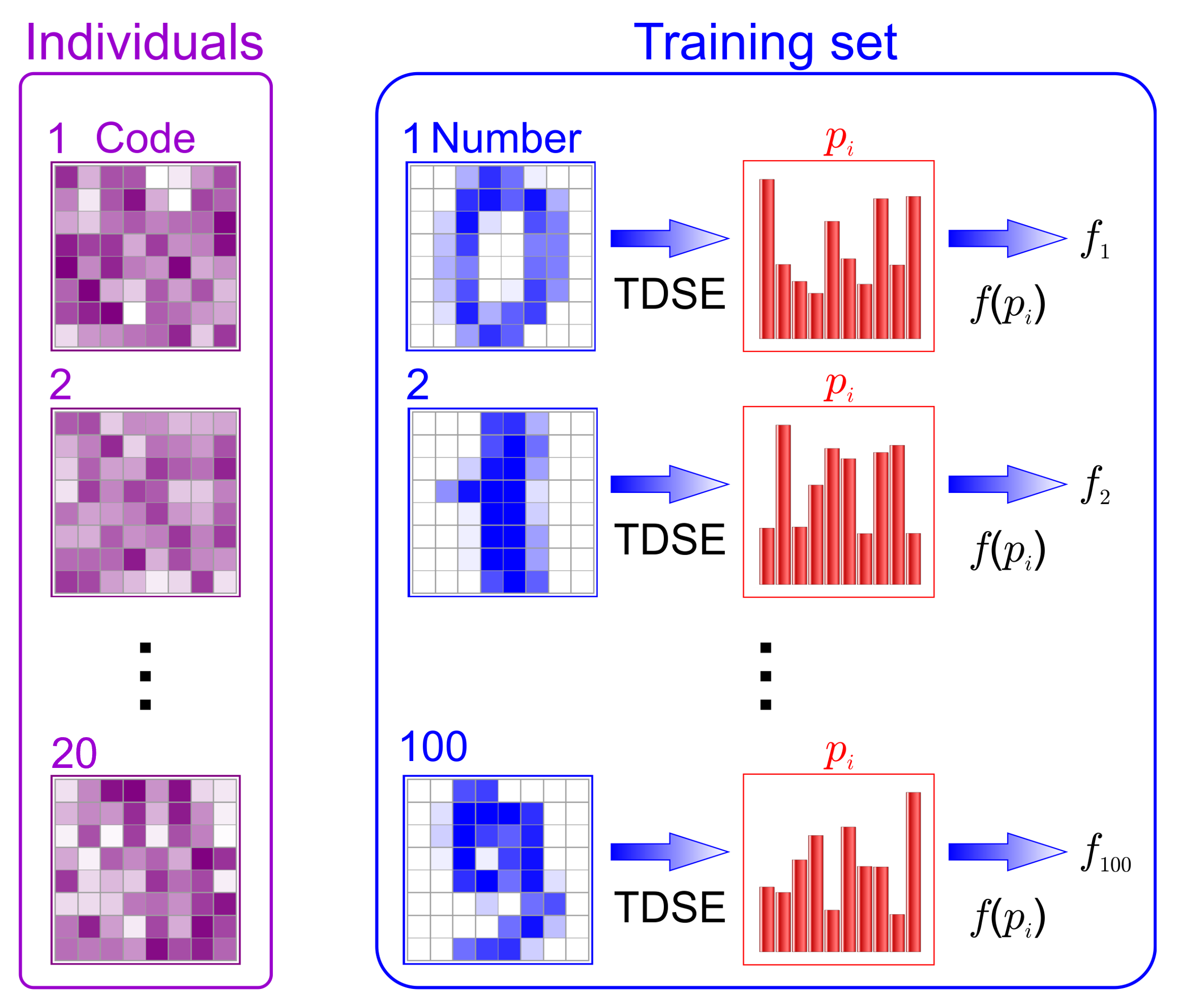}
    \caption{\label{fig:Pulses-Levels-Iterations} 
    Scheme for the iterative optimization of the code.  Each of the 20 individuals (represented by their 8x8 numerical codes) of one generation are put together with all 100 hand-written digits of the training set and each individual's fitness $F$ is obtained by the sum over all the single fitness values, $F=\sum_{n=1}^{100} f_n$.}
\end{figure}

In our simulations, we employ a multi-level Hamiltonian $\mathcal H_0$ to describe the atom, which is dipole coupled to an external arbitrarily polarized time-dependent laser field vector $\bm{E}(t)=E_x(t)\,\bm{e}_x+E_y(t)\,\bm{e}_y$
with Cartesian unit vectors $\bm{e}_x, \bm{e}_y$. 
The interacting Hamiltonian thus reads
\begin{equation}
    \mathcal H= \mathcal H_0 +  \mathcal V_x\,E_x(t) + \mathcal V_y\,E_y(t).
\end{equation}
Using a 20-level model with basis states 
1s, 2s, 2p$_{-1,1}$, 3s, 3p$_{-1,1}$, 3d$_{-2,0,2}$, 4s, 4p$_{-1,1}$, 4d$_{-2,0,2}$, 4f$_{-3,-1,1,3}$, we have
\begin{align}
&
    \mathcal H_0=
    \begin{pmatrix}
E_{\mathrm{1s}} & 0 & 0 & \\
0 & \!\!E_{\mathrm{2s}} & 0 & \\
0 & 0 & \!\!E_{\mathrm{2p}} & \\[-.23cm]
&&&\hspace*{-.25cm}{\mbox{\footnotesize{$\ddots$}}}
\end{pmatrix},
\end{align}
\begin{align}
    \label{coupling}    
    \mathcal V_x=
    \begin{pmatrix}
0 & 0 & a & -a &\\
0 & 0 & a & -a &\\
a & a & 0 & 0 &\\
-a & -a & 0 & 0 & \\[-.25cm]
&&&&\hspace*{-.22cm}{\mbox{\footnotesize{$\ddots$}}}
\end{pmatrix},
\end{align}
and
\begin{align}
\hspace{.2cm}
    \mathcal V_y=
    \begin{pmatrix}
0 & 0 & -\I a & -\I a \\
0 & 0 & -\I a & -\I a \\
\I a & \I a & 0 & 0 \\
\I a & \I a & 0 & 0 \\[-.25cm]
&&&&\hspace*{-.22cm}{\mbox{\footnotesize{$\ddots$}}}
\end{pmatrix}.
\end{align}
For simplicity, the energies are chosen as $E_{nl}=-1/(n+1)^2$. The coupling
matrix elements read
$\mathcal V_x^{jk} = \, \langle l,m | \sin\theta\cos\phi | l',m' \rangle$ and $\mathcal V_y^{jk} = \, \langle l,m | \sin\theta\sin\phi | l',m' \rangle$,
where the index $j$ corresponds to the state $|n,l,m\rangle$, the index $k$ corresponds to the state $|n',l',m'\rangle$, and $|l,m\rangle$
are angular momentum states. Here, the radial integrals have been set to unity to emphasize the generic character of the model. This means that, for example, the number $a$ appearing in Eq.~(\ref{coupling}) is $a = \langle 0,0|\sin\theta\cos\phi|1,-1\rangle$. The 1s state $|1,0,0\rangle$ is taken as the initial state for the time evolution.  

\revisedtext{
The orthogonally-polarized shaped laser fields~[27] $E_x$ and $E_y$ encode the input data (handwritten digits) in the $x$-component and the program data in the $y$-component. Each digit image and each program consists of 64 values. Essentially, we translate these values into the phases of 64 different frequencies, from which a pulse is composed. In detail, the encoding is implemented as follows. We start from a $\cos^2$-shaped spectral amplitude}
\begin{equation}
   \tilde{E}(\omega) = \tilde{E}_0
   \, \cos^2{
   \left( \frac{\omega-\omega_0}{\Omega}{\pi}  \right)},
\end{equation}
\begin{equation}
   \omega_0-\Omega/2 \leq\omega\leq\omega_0+\Omega/2,
\end{equation}
\revisedtext{
where $\omega$ is the frequency, the spectral range is $\Omega=\pi/32\, \rm a.u.$ and the central frequency $\omega_0$ corresponds to a laser wavelength of 800$\,$nm.
The time-dependent field $E(t)$ is then obtained as a Fourier-synthesis of 64 components, multiplied by an additional $\cos^2$ temporal envelope to restrict the pulse to a finite total duration $T$, i.e.,}
\begin{equation}
E(t) = \cos^2(t\pi/T) \, \sum_{j=0}^{63}\,\tilde{E}(\omega_j)\,
\cos(-\omega_j t + \varphi_j),
\end{equation}
\begin{equation}
-T/2\leq t \leq T/2
\end{equation}
with
\begin{align}    
    &\omega_j = \omega_0 + (j-31)\,\Delta\omega,\quad   
    \Delta\omega = 2\pi/T,\\
    &T=4096\,\textrm{a.u.},\\ 
    &\varphi_j = - 3v_j \textrm{rad}.
\end{align}
\revisedtext{
Here, $v_j$ (in the range $0\!\leq\!v_j \!\leq\! 1$) are the 64 data values describing either the input or the program. In the case of the input, $v_j$ are the pixel intensities of hand-written digits from the scikit-learn python package~[28], 8x8 pixel representation, processed column-wise from the lower right to the upper left corner. Examples of this parameterization are shown in  Fig. 2.
}

The TDSE is then solved numerically by a split-step operator approach \cite{Feit1982} with a time step of 1 a.u., and the final state populations $p_i$ (``output'') are read out for each pair of 100 input and 20 program fields. 
Among the excited states, we select 10 classification states (namely, all dipole-accessible 4s, 4p, 4d, 4f states) to represent the digits from ``0'' to ``9''.
Because not all classification states will be equally populated for an arbitrary set of pulses, we renormalize the final populations of the classification states as 
\revisedtext{$P_j=p_j/p_{0j}$ ($j=0,\dots,9$), 
where $p_{0j}$} is obtained by applying a set of random pc pulses and reading out their final state populations.  
\revisedtext{
This renormalization strategy makes the scheme robust against the particular choice of quantum system.
}
\revisedtext{
Any experimental implementation of the scheme needs to be performed on an ensemble of atoms, 
in order to perform not just a single measurement (determining the state $| n,l,m \rangle$ of a single atom), but to obtain the ensemble average, i.e., the probabilities of finding an atom in any of the target states.
The classification state with the highest renormalized population is then identified as the classification output between ``0'' and ``9''.
}

We assign to each program field a fitness $F=3N+P$ with $N$ the number of matches between the input (hand-written digit) and classification output (state with highest renormalized population).
\revisedtext{
The term $P$ is included to reward those pulses leading to particularly high populations of the correct classification states: $P=\sum_{n=1}^{100}P_{j(n)}^{(n)}$ is the sum over the renormalized populations $P_{j(n)}^{(n)}$ for each of the 100 input fields of the training set, with $j(n)$ being the correct classification state for the $n$th input field.
}
The program fields
are then iteratively optimized by an evolutionary algorithm to maximize their fitness. The goal is to find an optimal program field that guarantees correct classification of all training digits. The evolutionary algorithm employed here to train the model atom uses a population size of 20 individuals, each represented by an array of 64 numbers (its ``genes''), which determine the spectral phase of the pc pulse in the same way as for the input pulses. For each individual pc pulse its fitness is calculated by applying the pulse to 100 hand-written digits from the training set. The best pc pulse is always kept for the next generation (``cloning'') while the remaining 19 individuals are obtained by a combination of cross-over and mutation (using random numbers) for its 64-number genetic array.

\begin{figure}
    \begin{center}
    \includegraphics[width=\columnwidth]{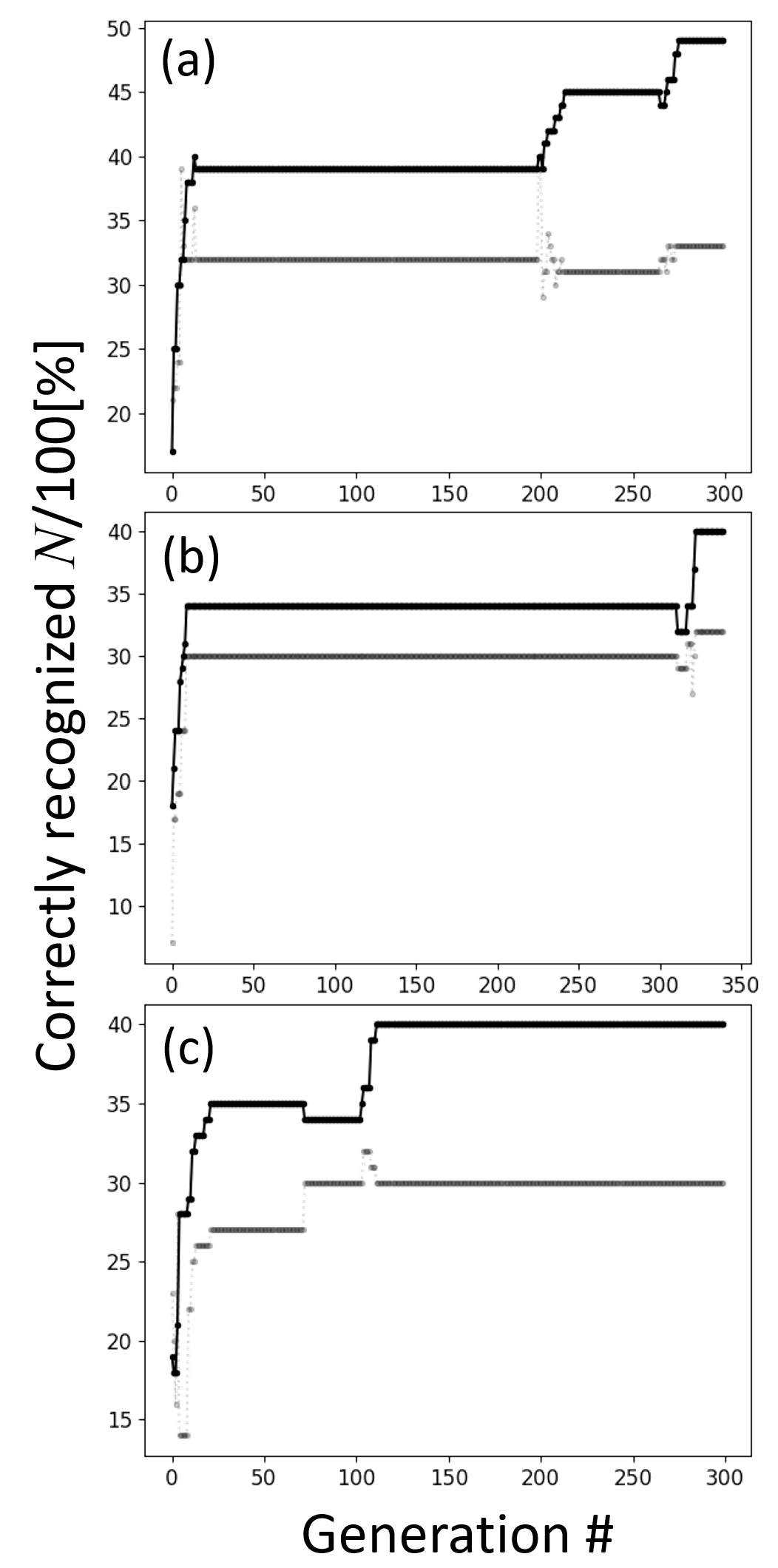}
    \end{center}
    \caption{\label{optimization} 
    \revisedtext{
    Percentage of correctly recognized digits for three example runs of the evolutionary algorithm to optimize the program fields for digit recognition.  (a) Model with state-independent couplings $a$ for all transitions, (b) model with state-dependent couplings $a'=a/(|n-n'|+1)$, (c) model with modified state energies $E_{nl}=-2.16/(n+2)^2$.  Black points: fittest individual when applied to the training set. Grey points: fittest individual when applied to an (unseen) test set for validation of the generalization of learning. Routinely, success rates of $>\,$40\% and $\sim\,$30\% 
    are achieved for training and test sets, respectively.
    }
    %
    }
\end{figure}

The results of three sample optimization runs are shown in Fig.~\ref{optimization}.  While  the proportion of correctly classified digits generally rises throughout the optimization (black line), it is particularly interesting to observe the correlated increase of the success rate on the test set (gray line). Since the 100-sample test set of digits is unknown to the algorithm during training with the 100-sample training set, the increase in the success rate on the test-set suggests that generalization is achieved in this approach, i.e., discrepancies between different hand-written versions of the same digit do not prevent a correct classification.

In summary, we have introduced a novel concept of optically programmable learning using quantum states for classification. As a proof of principle, the concept was applied to hand-written digit recognition, implemented with a few-level model of an atom and a straightforward encoding of input/data and program/code by spectral phase functions of two orthogonally polarized femtosecond optical laser fields. Once the optimal program pulse is known, the digit recognition code runs on the femtosecond time scale, which is much faster than the processing time of any classical or quantum computer. The ultrafast time evolution minimizes the influence of environment-induced decoherence and makes the proposed scheme robust. We note that replacing the atom by a larger quantum system, such as a complex molecule, enormously increases the size of the Hilbert space while preserving the ultrafast time scale. 
\revisedtext{
We have chosen the digit-recognition task as it is routinely used in conventional machine learning and we run it on a simple quantum system for which the TDSE can be solved numerically.
For more complex tasks implemented on larger systems, the numerical solution of the TDSE becomes prohibitive so that the actual experiment becomes the method of choice. 
}

The key advantage of our approach is based on the versatility of possible applications using the same quantum system as computational kernel. For example, future applications could include other computational tasks, such as identifying letters, images, or prime numbers. In our approach, the quantum system acts as an optically reprogrammable general-purpose quantum processor. 
While here we only introduce and explore the key idea by means of a one-particle few-level model, future experimental implementations will involve few- or many-body dynamics to employ a much larger state space and thus higher effective number of coupled layers of states (neurons). A crucial point that distinguishes this scheme from existing quantum-computing approaches and platforms is the fact that (entangling) operations are not performed on spatially \emph{separated} entities but on \emph{compact} quantum systems of interacting particles. It is therefore neither possible nor necessary to implement traditional quantum gates. Instead, suitably-shaped structured light pulses are used to perform the operations required for the envisaged task.  
\revisedtext{The use of ultrafast polarization shaping to transfer the data and the code into an atomic-scale quantum system as well as the execution within the system proceeds on femtosecond time scales, implying unprecedented speed when compared to classical computers as well as quantum computers using trapped ions or superconducting qubits. Therefore, this approach may lead to new scientific as well as technological applications, for example in real-time classification. 
It has the potential to outpace other currently implemented machine-learning approaches, including the fastest optical on-chip neuromorphic systems and optical accelerators, by orders of magnitude.
}

Our scheme can be viewed as a form of machine learning with a microscopic quantum system. Similar to classical machine learning, there is not necessarily a simple explanation for how the trained machine makes its decisions. This ``lack of insight'' has not prevented conventional artificial intelligence from revolutionizing technology in recent years. We thus expect that ultrafast artificial intelligence on the atomic scale has the potential to exploit quantum mechanics for high-speed computational tasks in the future.

\bibliography{bibatomicai}

\end{document}